\documentclass[twocolumn,prd,showpacs,floatfix,%
superscriptaddress,preprintnumbers,nofootinbib]{revtex4}

\usepackage{epsfig}
\usepackage{amsmath,bm}

\begin{document}

\title{Supernova neutrino halo and the suppression of self-induced
flavor conversion}

\author{Srdjan Sarikas}
\affiliation{Max-Planck-Institut f\"ur Physik
(Werner-Heisenberg-Institut), F\"ohringer Ring 6, 80805 M\"unchen,
Germany}

\author{Irene Tamborra}
\affiliation{Max-Planck-Institut f\"ur Physik
(Werner-Heisenberg-Institut), F\"ohringer Ring 6, 80805 M\"unchen,
Germany}

\author{Georg Raffelt}
\affiliation{Max-Planck-Institut f\"ur Physik
(Werner-Heisenberg-Institut), F\"ohringer Ring 6, 80805 M\"unchen,
Germany}

\author{Lorenz H\"udepohl}
\affiliation{Max-Planck-Institut f\"ur Astrophysik,
Karl-Schwarzschild-Str.~1, 85748 Garching, Germany}

\author{Hans-Thomas Janka}
\affiliation{Max-Planck-Institut f\"ur Astrophysik,
Karl-Schwarzschild-Str.~1, 85748 Garching, Germany}

\date{4 April 2012, published 19 June 2012}

\begin{abstract}
Neutrinos streaming from a supernova core occasionally scatter
in the envelope, producing a small ``neutrino halo'' with a much
broader angle distribution than the primary flux originating
directly from the core. Cherry {\it et al.} have recently pointed
out that, during the accretion phase, the halo actually dominates
neutrino-neutrino refraction at distances exceeding some 100~km.
However, the multiangle matter effect (which increases if the angle
distribution is broader) still appears to suppress self-induced
flavor conversion during the accretion phase.
\end{abstract}

\preprint{MPP-2012-63}


\maketitle

\section{Introduction}                               \label{sec:intro}

Neutrino-neutrino refraction is responsible for the intriguing
effect of self-induced flavor conversion~\cite{Samuel:1993uw,
Duan:2005cp, Hannestad:2006nj, Raffelt:2011yb, Banerjee:2011fj} that
can occur in the neutrino flux streaming from a supernova (SN)
core~\cite{Duan:2010bg}. In this context, the angular neutrino
distribution plays a crucial role. The current-current structure of
low-energy weak interactions implies that the interaction energy
between two relativistic particles involves a factor
$(1-\cos\theta)$ where $\theta$ is their relative direction of
motion.

If the neutrino-emitting region of a supernova core (``neutrino
sphere'') has radius $R$, then at distances $r\gg R$ a typical
neutrino-neutrino angle is $\theta\sim R/r$ and $\langle
1-\cos\theta\rangle\propto(R/r)^2$. The geometric flux dilution
provides another factor $(R/r)^2$, leading to an overall $(R/r)^4$
decrease of the neutrino-neutrino interaction
energy~\cite{Duan:2010bg}.

In a recent paper, Cherry {\it et al.} pointed out that this
picture is not complete because neutrinos suffer residual collisions
on their way out~\cite{Cherry:2012zw}. Every layer of matter above
the neutrino sphere is a secondary source, producing a wide-angle
``halo'' for the forward-peaked primary flux. While the halo flux is
small, its broad angular distribution allows it to dominate the
neutrino-neutrino interaction energy.

\begin{figure}[b]
\includegraphics[width=0.74\columnwidth]{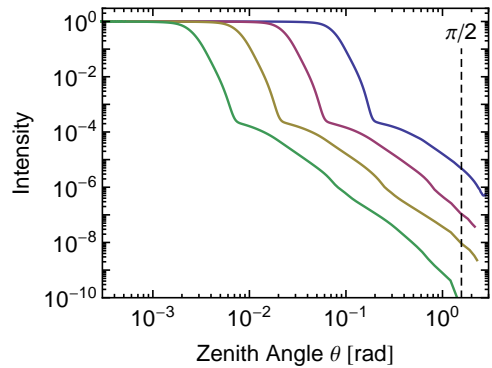}
\caption{Intensity for the $\bar\nu_e$ radiation field of our numerical model,
normalized to the
forward direction, measured at the radial distances 300, 1000,
3000, and 10,000~km (right to left).\label{fig:angle1}}
\end{figure}

We illustrate the halo with a numerical example, the 280~ms
postbounce snapshot of a spherical $15\,M_\odot$ model. We recently
used it as our benchmark case to study multiangle suppression of
self-induced flavor conversion~\cite{Sarikas:2011am}. In
Fig.~\ref{fig:angle1} we show the angular dependence of the
intensity\footnote{With ``intenstiy'' ${\cal I}$ we mean the
quantity ``neutrinos per unit area per unit time per unit solid
angle,'' integrated over the energy spectrum.} for the $\bar\nu_e$
radiation field, normalized to the forward direction, measured at
the radial distances 300, 1000, 3000, and 10,000~km. (The angular
distributions become noisy for $\theta\agt\pi/2$, where they are
currently not well provided by our simulations.) The core and halo
fluxes are two distinct components, the latter so small that it is
not visible on a linear plot. If we use $\theta_{\rm c}\sim0.1$ as
the edge of the core distribution for the 300~km case, we infer a
radius of $R\sim30$~km for the region where neutrinos begin to
stream almost freely. At larger distances, the angular scales are
squeezed by a factor $R/r$.

The impact on neutrino-neutrino refraction is illustrated by the
weak potential felt by a radially moving neutrino. In terms of the
zenith angle $\theta$ of the intensity ${\cal I}(\theta)$ we need
the quantity
\begin{equation}\label{eq:Adef}
\langle 1-\cos\theta\rangle =
\frac{\int_0^\pi d\theta\,\sin\theta\,(1-\cos\theta)\,{\cal I}(\theta)}
{\int_0^\pi d\theta\,\sin\theta\,{\cal I}(\theta)}\,.
\end{equation}
For small $\theta$, the integrand in the numerator expands as
$(1-\cos\theta)\sin\theta=\theta^3/2$, allowing the halo to
contribute significantly at larger distances where the distributions
are more squeezed. At 1000~km, for example, $\langle
1-\cos\theta\rangle$ is almost a factor of 10 larger than it would
be without the halo, in agreement with Cherry {\it et al.}~\cite{Cherry:2012zw}. The halo parts of the functions in
Fig.~\ref{fig:angle1} decrease roughly as $\theta^{-3}$ and we find
this implies
that $\langle 1-\cos\theta\rangle$ decreases roughly as $r^{-1}$
instead of $r^{-2}$.
The neutrino-neutrino interaction energy then
decreases roughly as $r^{-3}$ instead of $r^{-4}$.

The halo is important during the accretion phase when there is enough
matter for the primary flux to scatter on \cite{Cherry:2012zw}.
However, the same high matter density tends to suppress the
self-induced flavor instability~\cite{EstebanPretel:2008ni}. During
the early accretion phase, self-induced flavor conversions were found
to be typically suppressed~\cite{Sarikas:2011am, Chakraborty:2011gd}.
Does the halo change these conclusions? Its importance derives from
its broad angle distribution, which also makes it susceptible to the
multiangle matter suppression.

To investigate this question, we study in Sec.~\ref{sec:halo} the
properties of the neutrino halo. In Sec.~\ref{sec:stability} we
perform a stability analysis along the lines of our recent study of
an accretion-phase model~\cite{Sarikas:2011am}, and we conclude in
Sec.~\ref{sec:conclusions}.

\section{Neutrino Halo}                              \label{sec:halo}

\subsection{Numerical angle distribution}

Neutrinos stream almost freely and therefore, at larger distances,
the angular distributions are simply squeezed to smaller angular
scales (Fig.~\ref{fig:angle1}). This behavior applies also to the
halo flux which, once produced, streams almost freely. At larger
distances, the halo distribution primarily gains at the edges: the
newly available angular modes get populated by scattering.

\begin{figure}[b]
\includegraphics[width=0.74\columnwidth]{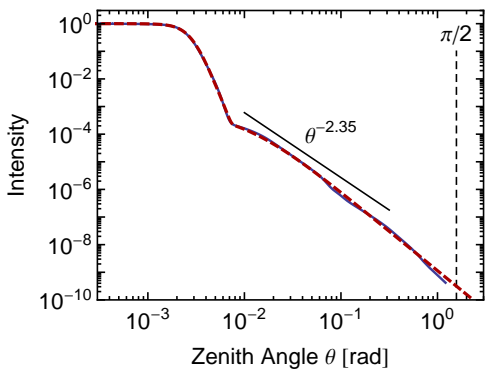}
\caption{The $10^4$~km numerical case
(solid line) overlaid with the analytic fit (dashed line) of
Eq.~(\ref{eq:fit}). We also indicate the power-law behavior estimated
from an analytic argument (Sec.~\ref{sec:analytichalo}).
\label{fig:angle3}}
\end{figure}

The numerical angular distribution at the largest available radius
essentially holds all the required information. The decreasing wings
of ${\cal I}(\theta)$ look like power laws and we can fit the entire
distribution by
\begin{eqnarray}\label{eq:fit}
{\cal I}_{\rm fit}(\theta)&=&\left[\left(\frac{0.9994}{[1+(\theta/0.0029)^{4.5}]^2}\right)^5\right.
\nonumber\\
&&\kern2em{}+\left.\left(\frac{0.0006}{[1+(\theta/0.01)^{1.43}]^2}\right)^5\right]^{1/5}\!\!.
\end{eqnarray}
We show this function overlaid with the $10^4$~km case in
Fig.~\ref{fig:angle3}. For the other species, the situation is
analogous with slightly different parameters.

\subsection{Energy distribution}
\label{sec:energydistribution}

\begin{figure}[t]
\includegraphics[width=0.74\columnwidth]{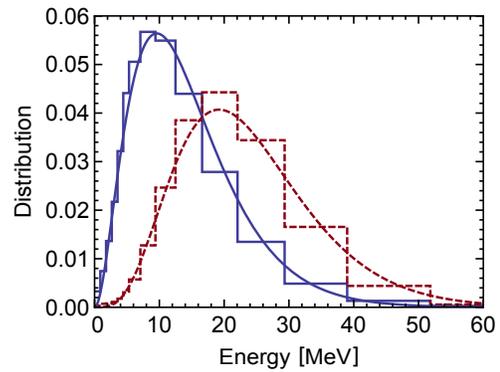}
\caption{Energy spectrum of core and halo components.
Histogram: Numerical results. Smooth lines: Thermal spectra as
described in the text with the same $T$.\label{fig:energy}}
\end{figure}

The $\bar\nu_e$ flux spectrum emitted from the core roughly follows
a thermal Maxwell-Boltzmann form $f_{\rm c}(E)\propto
E^2\,e^{-E/T}$. In Fig.~\ref{fig:energy} we show as a histogram
(blue) the numerical spectrum of the core component (measured in the
forward direction) together with a thermal fit with $T=4.8$~MeV
(average energy 14.4~MeV).

The halo component arises from scattering on nuclei with a cross
section proportional to $E^2$. Therefore, the halo spectrum should
be $f_{\rm h}(E)\propto E^4\,e^{-E/T}$ with the same $T$. In
Fig.~\ref{fig:energy} we show as a red histogram the numerical halo
spectrum (measured at a very large angle) together with such a fit.
In the spirit of an overall consistency check, we find excellent
agreement.

\subsection{Neutrino-neutrino refraction}

The impact on neutrino-neutrino refraction is quantified by $\langle
1-\cos\theta\rangle$ as defined in Eq.~(\ref{eq:Adef}). Motivated by
the numerical examples, we first consider truncated power-law
intensity distributions of the form
\begin{equation}
{\cal I}(\theta)=\begin{cases}1&\hbox{for $\theta\leq\theta_{\rm c}$}\\
(\theta_{\rm c}/\theta)^p&\hbox{otherwise.}\end{cases}
\end{equation}
We ask for the asymptotic behavior of $\langle 1-\cos\theta\rangle$
at large distance, corresponding to $\theta_{\rm c}\ll1$. The
integrand in Eq.~(\ref{eq:Adef}) expands as
$\sin\theta\,(1-\cos\theta)\to\theta^3/2$. If $p>4$, the integral is
dominated by small angles and we find the result shown in
Table~\ref{tab:powerlaws}. In other words, if ${\cal I}(\theta)$
falls off fast enough, we recover the classic $r^{-2}$
scaling, where we have used that $\theta_{\rm c}\sim R/r$.

For $p\leq4$ we can no longer extend the upper integration limit to
$\infty$ and can no longer expand the integrands. With Mathematica
we find analytic results with coefficients for $\theta_{\rm c}\ll 1$
that are given in Table~\ref{tab:powerlaws}. In particular, for
$p=3$ the scaling is linear in $\theta_{\rm c}$.

\begin{table}
\caption{Average $\langle 1-\cos\theta\rangle$ for $\theta_{\rm
c}\ll1$ and different $p$.\label{tab:powerlaws}}
\begin{ruledtabular}
\begin{tabular}{ll}
Power $p$&$\langle 1-\cos\theta\rangle$\\[0.4ex]
\hline
$\infty$&$\displaystyle\frac{1}{4}\,\theta_{\rm c}^2$\\[1ex]
$p>4$&$\displaystyle\frac{p-2}{p-4}\,\,\frac{\theta_{\rm c}^2}{4}$\\[1ex]
4&$\displaystyle(1.1897-2\log\theta_{\rm c})\,\frac{\theta_{\rm c}^2}{4}$\\[1ex]
3&$\displaystyle0.54033\,\theta_{\rm c}$\\[1ex]
2&$\displaystyle\frac{1.5788}{1.9929-2\log\theta_{\rm c}}$\\[1ex]
1&0.61712\\[0ex]
0&1\\
\end{tabular}
\end{ruledtabular}
\end{table}

\begin{figure}[b]
\includegraphics[width=0.74\columnwidth]{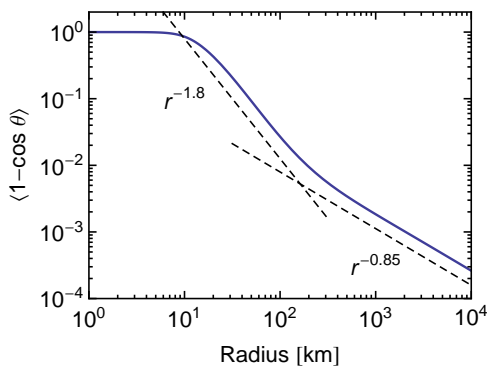}
\caption{Average $\langle 1-\cos\theta\rangle$ for our analytic
angle distribution of Eq.~(\ref{eq:fit}), representing the 280~ms numerical model.
\label{fig:average1}}
\end{figure}

\begin{figure}[t]
\includegraphics[width=0.74\columnwidth]{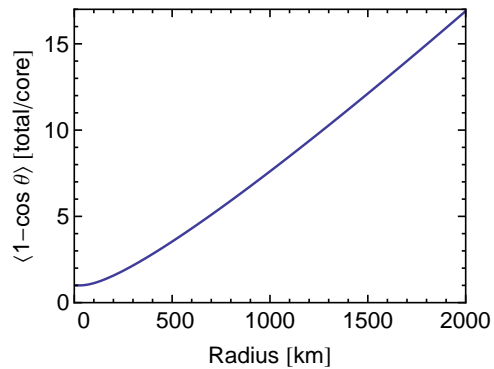}
\caption{Same as Fig.~\ref{fig:average1}, now showing the
enhancement caused by the halo, i.e.\ the ratio total/core
contribution, where the core flux is attributed to the first term in
Eq.~(\ref{eq:fit}). \label{fig:average2}}
\end{figure}

Next we consider the analytic fit of Eq.~(\ref{eq:fit}) and show
$\langle 1-\cos\theta\rangle$ in Fig.~\ref{fig:average1}. We
integrate up to $\theta=\pi$ for each radius, which means that we smoothly interpolate between small and large distances. At
small radii (inside of the SN core) the distribution is isotropic
and $\langle 1-\cos\theta\rangle=1$. For $r\alt200$~km we find
approximately the naive $r^{-2}$ scaling. At larger distances, the
halo becomes important and the scaling turns approximately to
$r^{-1}$. If the halo distribution would scale as $\theta^{-3}$ we
would expect precisely $r^{-1}$, but our fit corresponds to ${\cal
I}(\theta)\propto\theta^{-2.86}$, explaining the less steep
variation.

In Fig.~\ref{fig:average2} we show the enhancement of $\langle
1-\cos\theta\rangle$ caused by the halo, where we attribute the
first part of the analytic fit of Eq.~(\ref{eq:fit}) to the core.
The enhancement caused by the halo scales almost linearly at large
distances. At $r\sim 1000$~km the enhancement is about a factor
of~8, roughly in agreement with Cherry {\it et al.}\ \cite{Cherry:2012zw}.

\subsection{Analytic halo estimate} \label{sec:analytichalo}

For an analytic estimate, we consider a total neutrino production
rate $\Phi=L/\langle E\rangle$ emitted from a pointlike source
(neutrino luminosity $L$). It traverses a spherical matter
distribution which we model as a decreasing power law of the form
$n(r)=n_R(R/r)^m$. We assume that multiple scatterings can be
neglected. Every spherical shell is a neutrino source from
scattering the primary flux. For the scattering cross section, we
assume the form $d\sigma/d\Omega=\sigma_{\rm
scatt}(1+a\cos\vartheta)/4\pi$, where $\vartheta$ is the scattering
angle. Elementary geometry (see the Appendix) reveals that at distance $r$ we expect a
scattering flux (halo) of
\begin{eqnarray}\label{eq:halo-analytic}
{\cal I}(r,\theta)&=&\frac{\Phi\,\sigma_{\rm scatt}n_R}{(4\pi)^2R}\,
\left(\frac{R}{r\,\sin\theta}\right)^{m+1}\nonumber\\
&&{}\times
\int_\theta^\pi d\vartheta\,(1+a\cos\vartheta)\,\sin^m \vartheta\,.
\end{eqnarray}
For small angles, we find a power law ${\cal
I}(r,\theta)\propto\theta^{-(m+1)}$, whereas the large-angle and
backward flux depend on the detailed cross-section angular
dependence.

In our numerical example, the density outside the shock wave (at
about 70~km) decreases roughly as $r^{-1.35}$ out to about 5000~km.
With $m=1.35$ we expect the halo to vary as $\theta^{-2.35}$ as
indicated in Fig.~\ref{fig:angle3}. While it would not provide a
good global fit, it looks excellent for about the first decade of
angles of the halo.

Based on the analytic expression for ${\cal I}(r,\theta)$ we can
estimate the inward-going flux. It is very small compared to the
outward-going core flux, but its contribution to refraction need not
be small. We consider a region at sufficiently large radius where
neutrino-neutrino refraction is dominated by the halo and ask which
fraction is caused by inward-going neutrinos, i.e.\ which fraction of
the integral in Eq.~(\ref{eq:Adef}) is contributed by
$\pi/2<\theta\leq\pi$. For sufficiently small power-law index $m$
this fraction does not depend on the radius $r$ (to lowest order).
Coherent neutrino-nucleus scattering is forward peaked and we use
$a=1$. For $m=1$ the contribution of the backward flux is 25\%, for
$m=2$ approximately 16\%.

\section{Stability Analysis}                     \label{sec:stability}

All previous studies of collective flavor oscillations of SN
neutrinos used the free-streaming approximation: it was assumed that
collisions play no role in the relevant stellar region. On the other
hand, the halo dominates neutrino-neutrino refraction at larger
radii, so in some sense the free-streaming assumption gets worse
with distance, not better. Therefore, in order to understand
collective flavor conversion in this region, one may have to
rethink the overall approach~\cite{Cherry:2012zw}.

On the other hand, the multiangle matter effect may still suppress
the onset of collective flavor conversions, at least in some
accretion-phase models. A linearized stability analysis would still
be possible and self-consistent. If self-induced flavor conversions
have not occurred up to some radius and if the matter effect is
large in that region, then neutrinos continue to be in
weak-interaction eigenstates up to small corrections caused by
oscillations with the small in-matter effective mixing angle.
Therefore, we can use the core+halo flux at that radius as an inner
boundary condition for the subsequent evolution. In other words,
collisions at smaller radii do not change the conceptual approach,
they only provide a broader angular distribution than would have
existed otherwise.

A greater problem is how to deal with the inward-going flux which,
for our conditions, contributes some 20\% to neutrino-neutrino
refraction. We will simply neglect it because it cannot be
incorporated self-consistently in an approach based on outward-only
streaming neutrinos. If the system is stable and stationary, then the
picture remains self-consistent because neutrinos remain in flavor
eigenstates, including the backward-moving ones.

In this spirit we repeat our linearized flavor stability analysis for
our 280~ms snapshot of a $15\,M_\odot$ accretion-phase
model~\cite{Sarikas:2011am} in a simplified form. We model the
angular distribution according to our analytic fit,
Eq.~(\ref{eq:fit}), which we use at all distances in the form
\begin{equation}
{\cal I}(r,\theta)\propto {\cal I}_{\rm fit}(\theta\,r/10^4~{\rm km})\,,
\end{equation}
because the curves shown in Fig.~\ref{fig:angle1} are almost
identical up to an angle scaling. We cut the distribution at
$\theta=\pi/2$ to avoid backward-going modes. We assume
monoenergetic neutrinos with $\omega_{\rm c}=\Delta m^2/2E_{\rm
c}=0.5~{\rm km}^{-1}$ for the core flux, i.e.\ $E_{\rm c}\sim
12$~MeV. We have checked that this choice reproduces well the
instability region in Fig.~4 of our previous
paper~\cite{Sarikas:2011am} where the full spectrum was used. The
halo energies are larger as described in
Sec.~\ref{sec:energydistribution}, implying $\langle E^{-1}_{\rm
c}\rangle=2\langle E^{-1}_{\rm h}\rangle$, thus we use monoenergetic
halo neutrinos with $\omega_{\rm h}=0.25~{\rm km}^{-1}$.

\begin{figure}[t]
\includegraphics[width=0.85\columnwidth]{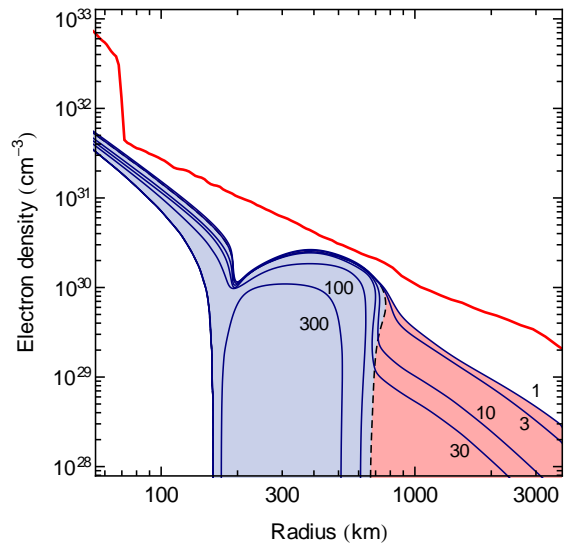}
\caption{Contours for the indicated $\kappa r$ values in the parameter
  space of radius $r$ and assumed electron density $n_e$. We use a
  simplified energy spectrum as described in the text. The core flux
  alone is responsible for $\kappa r\gg 1$ for $r\alt 700$~km (blue
  shading), the halo adds the red-shaded region at larger $r$. We also
  show the density profile of our 280~ms numerical
  model.\label{fig:is_regions}}
\end{figure}

The system is unstable in flavor space if the linearized equation of
motion discussed in Ref.~\cite{Sarikas:2011am} has eigenvalues with
imaginary part $\kappa$, the radial growth rate (in units of
km$^{-1}$) of the unstable mode. Whether this growth rate is large or
small depends on the available distance. In other words, at a given
distance $r$, the instability is important if $\kappa r\gg 1$ and
unimportant if $\kappa r\ll 1$. Therefore, in
Fig.~\ref{fig:is_regions} we show contours of constant $\kappa r$ in
a two-dimensional parameter space consisting of radius $r$ of our
model and an assumed electron density $n_e$ which causes the
multiangle matter effect.

If we use the core flux alone, this figure contains the same
information as Fig.~4 in our previous paper~\cite{Sarikas:2011am}. In
the absence of matter, the instability would set in at about 150~km,
whereas the presence of matter suppresses the instability entirely:
the SN density profile never overlaps with the instability region.
The most dangerous region of closest approach is at around 600~km.

In Fig.~\ref{fig:is_regions} one clearly sees how the halo flux
extends the instability region to larger radii (red shading), but
only far away from the SN density profile. In other words, the
multiangle matter effect strongly suppresses the would-be
instability caused by the halo.

This finding does not contradict the importance of the halo flux for
neutrino-neutrino refraction. It is true that the neutrino-neutrino
interaction energy caused by the halo
was found to decrease roughly as $r^{-3}$,
instead of $r^{-4}$ by the core flux alone. However,
the multiangle matter effect is enhanced in a similar way because it
depends on the same neutrino angular distribution. A more appropriate
comparison is between the neutrino and electron densities.
The former decreases as $r^{-2}$ whereas in our
model the latter decreases roughly as $r^{-1.35}$ and thus
becomes relatively more important with distance. Therefore, it is no
surprise that the multiangle matter effect would be more important
with increasing distance.

\section{Conclusions}                          \label{sec:conclusions}

The halo distribution of neutrinos causes neutrino-neutrino
refraction to decrease more slowly with distance than had been
thought, but still faster than the matter density during the
accretion phase. Repeating our linearized stability analysis
including the halo, the system remains stable for the chosen example.
As anticipated, the multiangle matter effect is a crucial ingredient
for collective flavor conversions especially in those regions where
the halo flux is important.

Thus it remains possible that self-induced flavor conversions are
generically suppressed during the early accretion phase. The early
signal rise would then encode a measurable imprint of the neutrino
mass hierarchy~\cite{Chakraborty:2011ir}.

However, more systematic studies are required to understand if such
conclusions are generic. For example, we have assumed that the
relatively small backward flux can be neglected for the stability
analysis, but its possible role needs to be better understood.
Likewise, the role of multi-D effects in SN modeling remains to be
explored.

The topic of collective flavor oscillations of SN neutrinos remains a
fiendishly complicated problem.

\section*{Acknowledgements} 

This work was partly supported by the Deutsche
Forschungsgemeinschaft under Grants No. TR-7 and No. EXC-153,
and by the  European Union FP7 ITN INVISIBLES (Marie Curie Actions, PITN-GA-2011-289442).
I.T.\ thanks the Alexander von Humboldt Foundation for support.

\begin{figure}[h!]
\includegraphics[width=0.70\columnwidth]{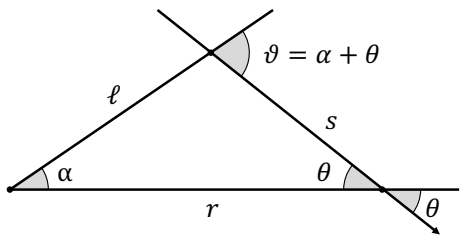}
\caption{Geometric definitions for the analytic halo
estimate.\label{fig:geometry}}
\end{figure}

\appendix*\section{Analytic Halo Estimate}         \label{sec:appendix}

In order to derive the analytic halo estimate of
Eq.~(\ref{eq:halo-analytic}) we consider a radial distribution of
scattering targets, $n(\ell)$, and a differential scattering cross
section $d\sigma/d\Omega$. The scattering rate per volume and per
solid angle for neutrinos with number density $n_{\nu}(\ell)$ at
radius $\ell$ is
\begin{equation}\label{eq:scattrate}
\frac{d q_\mathrm{scatt}}{d \Omega} =
n_{\nu}(\ell)\,n(\ell)\,c\,\frac{d \sigma}{d \Omega} \,,
\end{equation}
where we have reinstated the speed of light. The ``number
intensity'' ${\cal I}(r,\theta)$ at radius $r$ of neutrinos scattered
to angle direction $\theta$ relative to the radius vector at $r$ can
be obtained by adding up all scattering contributions along the path
$s$ as shown in Fig.~\ref{fig:geometry}. It depends on the total
neutrino emission rate $\Phi = L/\left\langle E\right\rangle$ of the
pointlike source assumed to be located at $r = \ell = 0$, which
determines the neutrino number density $n_{\nu}(\ell) = \Phi/(4\pi
\ell^2 c)$. With this result and Eq.~(\ref{eq:scattrate}) one finds
\begin{eqnarray}
{\cal I}(r,\theta) &=& \int_0^\infty ds\,
\frac{d q_\mathrm{scatt}}{d \Omega}(\alpha+\theta)  \nonumber \\
&=& \int_0^\infty ds\,
\frac{\Phi}{4\pi \ell^2}\,n(\ell)\,
\frac{d\sigma}{d\Omega}(\alpha+\theta)\,,
\label{eq:iintegral}
\end{eqnarray}
where $\vartheta=\alpha+\theta$ is the scattering angle and 
$\ell = r\sin\theta/\sin \vartheta$ and 
   $s = r\sin\alpha/\sin \vartheta$, see Fig.~\ref{fig:geometry}.
   
Using a decreasing power-law distribution of the target density,
$n(\ell) = n_R(R/\ell)^m$ with $m \ge 0$, an assumed dependence of
the scattering cross section on the scattering angle $\vartheta$ of
the form $d\sigma(\vartheta)/d\Omega = \sigma_\mathrm{scatt}(1 +
a\cos\vartheta)/4\pi$, one can transform the integral of Eq.~(\ref{eq:iintegral}) into the form of Eq.~(4).


\end{document}